\documentclass[prd,superscriptaddress,amsmath,twocolumn]{revtex4-2}
\usepackage{graphicx}
\usepackage{float}
 \pdfoutput=1
\usepackage{epstopdf,cancel}
\usepackage{epsf,latexsym,bbm,euscript}
\usepackage{amssymb,amsmath}
\usepackage{mathtools} 
\usepackage{times,graphics}
\usepackage{soul,xcolor}
\usepackage{mathtools}
\usepackage{mathrsfs}

\usepackage[caption=false]{subfig}
\newcommand{\subfigimg}[3][,]{%
  \setbox1=\hbox{\includegraphics[#1]{#3}}
  \leavevmode\rlap{\usebox1}
  \rlap{\hspace*{-10pt}\raisebox{.5\baselineskip}{\small{#2}}}
  \phantom{\usebox1}
}

\def\6{{\langle}}
\def\9{{\rangle}}
\newcommand{\defeq}{\vcentcolon=}
\newcommand{\eqdef}{=\vcentcolon}

\newcommand{\be}{\begin{equation}}
\newcommand{\ee}{\end{equation}}
\newcommand{\ba}{\begin{eqnarray}}
\newcommand{\ea}{\end{eqnarray}}

\def\sg{\textsl{g}}

\def\rf{\mathrm{f}}

\def\rin{\mathrm{in}}

\def\vS{v_\mathrm{f}}

 \newcommand{\mA}{{\mathrm{A}}}

\def\half{{\tfrac{1}{2}}}

\def\pad{{\partial}}

\def\sg{\textsl{g}}

\def\cO{\mathcal{O}}

\usepackage{url,hyperref}
\hypersetup{colorlinks,linkcolor={blue!55!black},citecolor={red!50!black},urlcolor={blue!45!black},breaklinks=true}

\begin{document}

\title{Matter and forces near physical black holes}

\author {Pravin K. Dahal}
\email{pravin-kumar.dahal@hdr.mq.edu.au}

\author{Ioannis Soranidis}
\email{ioannis.soranidis@hdr.mq.edu.au}

\author{Daniel R.\ Terno}
\email{daniel.terno@mq.edu.au}
\affiliation{School of Mathematical and Physical Sciences, Macquarie University, Sydney, New South Wales 2109, Australia}

\begin{abstract}

We describe   general features of formation and disappearance of regular spherically symmetric black holes in semiclassical gravity. The allowed models are critically dependent on the requirement that the resulting objects evolve in finite time according to a distant observer. Violation of the null energy condition (NEC) is mandatory for this to happen, and we study the properties of the necessary energy-momentum tensor in the vicinity of the apparent horizon. In studies of the kinematics of massive test particles, it is found that the escape from a black hole is possible only on the ingoing trajectories when the particles are overtaken by the contracting outer apparent horizon. Tidal forces experienced by geodesic observers, infalling or escaping, are shown to be finite at the apparent horizon, although this is not true for nongeodesic trajectories.

\end{abstract}
\maketitle

\section{Introduction}

Black holes are particularly elegant solutions of the Einstein equations. They introduce nontrivial causal structure into spacetime \cite{HE:73,MTW:73,FN:98,P:04}. About 100 ultracompact objects are identified as astrophysical black holes \cite{LIGO:21}. Black holes are domains of strong gravity, arguably the ones that are most accessible to observation. They may exemplify the conceptual tension between quantum mechanics and general relativity. They also may provide some clues about quantum gravity. Given all these different roles, it is still unclear if they are played by the one and the same actor. In fact, the variety of definitions of  black holes     matches this diversity of the roles~\cite{eC:19}.

It is useful to adapt the terminology of Ref. \cite{F:14} that distinguished between the mathematical and physical black holes. A mathematical black hole is a solution of the Einstein equations of classical
general relativity. It is the source of our ideas about what are the typical black hole features. The most well known of them is the event horizon, which for the Schwarzschild black hole is located
at the gravitational radius $r_\sg =2GM/c^2$. It separates
an interior spacetime containing a singularity from the outside observers.

All current observational
data can be explained within this paradigm. However,
an event horizon is a global teleological construct and is not accessible to
local observers \cite{H:00,mV:14}. On the other
hand, a trapped spacetime region from which currently nothing, not even light, can escape
—a crucial black hole property --- constitutes what one would reasonably regard as
a physical black hole (PBH). A trapped
region is a domain where both ingoing and outgoing future-directed
null geodesics emanating from a spacelike two-dimensional
surface with spherical topology have negative
expansion \cite{HE:73,vF:15}. The apparent horizon is the outer
boundary of the trapped region (here and elsewhere, we use the same name for both the 2D entity on a particular time slice and its 3D development; Ref~\cite{MMT:22} collects various relevant definitions).

It turns out that a careful analysis of the consequences of this definition, together with two natural assumptions (that we describe below in  Sec.~\ref{pbh}), provide a strong constraint on the near-horizon behavior of the possible models \cite{MMT:22}. In this work, we extend the previous results to identify several black hole properties that are important both for resolving conceptual issues and for modeling ultracompact objects.

We assume the validity of semiclassical gravity. That means
we use classical notions such as horizons and consider test particles with well-defined trajectories. The semiclassical Einstein equations

\be
G_{\mu\nu}\defeq R_{\mu\nu}-\half \sg_{\mu\nu} R=\6\hat T_{\mu\nu}\9_\omega\eqdef T_{\mu\nu}
\ee
describe the dynamics. Here the standard left-hand side is equated to the expectation
value of the renormalized energy-momentum
tensor (EMT). The latter represents both
the collapsing matter and the created excitations of the quantum fields.

Apart from assuming the validity of the semiclassical gravity, we make two further assumptions \cite{MMT:22}. First, we assume the weakest form of the cosmic
censorship conjecture. Usually, it is a statement
  that event horizons obscure spacetime singularities. Here we assume only that all curvature scalars that are
built from polynomials of components of the Riemann tensor are finite in some neighborhood of the apparent horizon \cite{BMMT:19}.

Second, we assume that the trapped region forms at a finite time of a distant observer (which we refer to as Bob) \cite{BMMT:19}. This is the only possible interpretation of regular black holes --- transient trapped regions without the event horizon and singularity. Formulation of the information loss problems requires both the Hawking radiation and the transient event horizon. Existence of the latter implies that the apparent horizon forms in finite time of Bob as well \cite{MMT:22D}.

The resulting analysis is based on self-consistency \cite{BMMT:19,MMT:22}. We study the semiclassical properties of the near-horizon geometry that follow from its existence that is subject to the two above assumptions. In particular, no global aspects of the spacetime structure, nature of the state $\omega$ or presence of the Hawking radiation are assumed.
We restrict the discussion to spherical symmetry. Because of its simplifying assumptions, the self-consistent approach results in a nearly complete description of the near-horizon geometry and physics.

\begin{figure*}[!htbp]
  \centering
  \begin{tabular}{@{\hspace*{0.025\linewidth}}p{0.45\linewidth}@{\hspace*{0.05\linewidth}}p{0.45\linewidth}@{}}
  	\centering
   	\subfigimg[scale=0.65]{(a)}{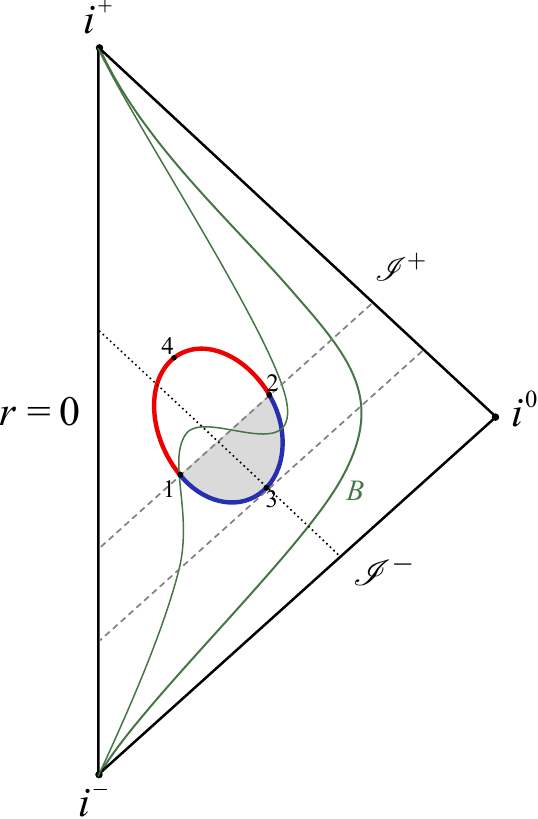} &
   	\subfigimg[scale=0.65]{(b)}{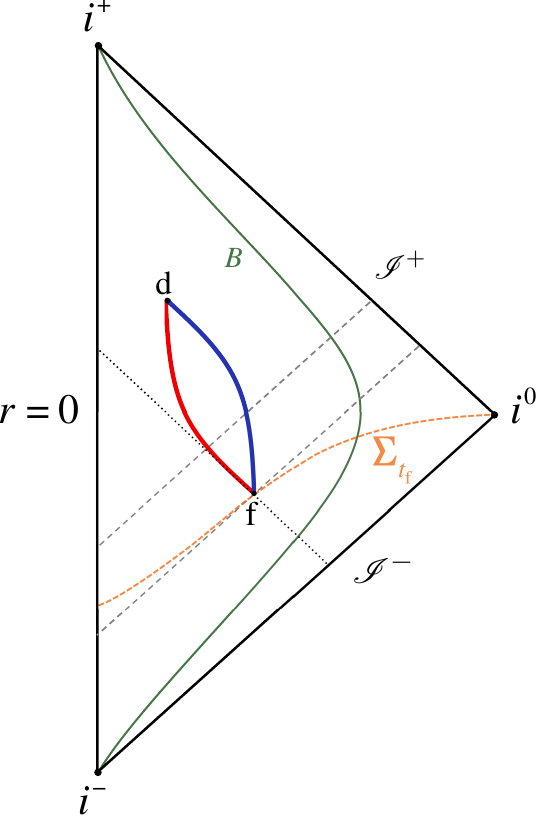}
  \end{tabular}
  	\caption{Schematic Carter–Penrose diagram for depicting formation and evaporation of a conventional RBH (a) and a  RBH that is treated as a PBH (b). The trajectory of a distant observer, Bob, is indicated in green and marked by the initial $B$. The dashed gray lines correspond to outgoing radial null geodesics that reach the future null infinity $\mathscr{I}^{+}$, and the dotted lines represent the ingoing radial null geodesics.  The asymptotic structure of a simple RBH spacetime coincides with that of Minkowski spacetime. An immediate neighborhood of $r = 0$ never belongs to the trapped region. (a) The outer (132, dark blue) and inner (142, dark red) apparent horizons are indicated according to the invariant definition of Eq.~\eqref{in-out}. These also correspond to the largest and the smallest roots of $f(v,r)=0$. This RBH has a smoothly joined inner and outer horizons \cite{BHL:18,CDLV:20}. The quantum ergosphere is indicated by the light gray shading. One of the hypersurfaces $r=\mathrm{const}$ is shown as a curved line that connects $i^-$ and $i^+$ and goes through the trapped region.  The NEC is satisfied along the segment (413). The segments (14) and (23) are timelike. (b) RBH treated as PBH with the outer (dark blue) and inner (dark red) apparent horizons. The points $\mathrm{f}$ and $\mathrm{d}$ represent the events of formation and disappearance of the trapped region. The equal time hypersurface $\Sigma_{t_{\mathrm{f}}}$ is shown as a dashed orange line connecting $r=0$ and $i^0$. The outer and the inner horizons are timelike (membranes). }
  	\label{fig:rbh}
\end{figure*}

The rest of this paper is organized as follows. First, we review the properties of PBHs in Sec.~\ref{pbh}. We also present the necessary conditions that any model of a regular black hole should satisfy. We discuss the classification of the EMT in Sec.~\ref{emt}. In Sec.~\ref{exit} we present some peculiar aspects of escaping massive test particles from a black hole. We consider the tidal forces experienced by an observer (Alice) in the vicinity of the apparent horizon in Sec.~\ref{tidal}.

We use the $(-+++)$ signature of the metric and work in units where $\hbar=c=G=1$.  Derivatives of a function of a single variable are marked with a prime:  $r'_g(t)\equiv dr_g/dt$, $r'_{+}(v)\equiv dr_{+}/dv$, etc. Derivatives with respect to the proper time $\tau$ or the affine parameter $\lambda$ are denoted by the dot: $\dot r=dr/d\tau$. We refer to a distant stationary observer as Bob and a traveling observer in the vicinity of the apparent horizon as Alice.

\section{Physical black holes}\label{pbh}

The self-consistent approach is best illustrated by the example of regular black holes (RBHs). Figure~\ref{fig:rbh}(a) is a sketch of a generic regular black hole, and Fig.~\ref{fig:rbh}(b) illustrates the features that necessarily arise when this putative object is treated as a physical black hole.

A general spherically symmetric metric in Schwarzschild coordinates is given by \cite{HE:73,vF:15}
\begin{align}
	ds^2 = -e^{2h(t,r)}f(t,r)dt^2+f(t,r)^{-1}dr^2+r^2d\Omega_2 , \label{metric}
\end{align}
where $r$ is the circumferential radius and $d\Omega_2$ is the area element on a unit two-sphere. These coordinates provide geometrically preferred foliations with respect to Kodama time \cite{vF:15}.
Some of the derivations become more transparent when they are expressed in radiative coordinates. Using the advanced null coordinate $v$, the metric is written as
\begin{align}
	ds^2=-e^{2h_+}f(v,r)dv^2+2e^{h_+}dvdr +r^2d\Omega_2 .  \label{metric-vr}
\end{align}
The    Misner-Sharp (MS) mass $C(t,r)/2\equiv M(t,r)$  is invariantly defined \cite{vF:15,FEFHM:17} via
\begin{align}
	\pad_\mu r\pad^\mu r\defeq f(t,r) \eqdef 1 - C/r , \label{MSmass}
\end{align}
and thus $C(t,r) \equiv C_+ \big( v(t,r),r \big)$.  The functions $h(t,r)$ and $h_+(v,r)$ play the role of integrating factors in coordinate transformations \cite{MMT:22}, such as
\begin{align}
	dt=e^{-h}(e^{h_+}dv- f^{-1}dr) . \label{intfactor}
\end{align}
For the Schwarzschild metric $C=2M=\mathrm{const}$ and $h\equiv 0$, while the coordinate $v$ becomes the ingoing   Eddington-Finkelstein coordinate.

The trapped region corresponds to the spacetime domain $f\leqslant 0$. The Schwarzschild radius $r_\sg(t)$ is the largest root of $f(t,r)=0$. Because of the invariance of $C$, it is invariant in the sense that $r_\sg(t) \equiv r_+\big(v(t,r_\sg))$.  Hence the outer apparent horizon  is located at the Schwarzschild radius $r_\sg$ \cite{vF:15,FEFHM:17}, justifying the definition of the black hole mass as  $2M(v)=r_+(v)$. Despite the fact that the apparent horizon is observer dependent in general, in spherically symmetric spacetimes, it is invariantly defined in all foliations that respect this symmetry \cite{FEFHM:17}.

It is convenient to introduce the effective EMT components
\begin{align}
	\tau{_t} \defeq e^{-2h} {T}_{tt}, \qquad {\tau}{^r} \defeq T^{rr}, \qquad  \tau {_t^r} \defeq e^{-h}  {T}{_t^r} . \label{eq:mtgEMTdecomp}
\end{align}
In spherical symmetry, the three Einstein equations (for the components $G_{tt}$, ${G}{_t^r}$, and ${G}^{rr}$) are
\begin{align}
	\partial_r C &= 8 \pi r^2  {\tau}{_t} / f , \label{eq:Gtt} \\
	\partial_t C &= 8 \pi r^2 e^h  {\tau}{_t^r} , \label{eq:Gtr} \\
	\partial_r h &= 4 \pi r \left(  {\tau}{_t} +  {\tau}{^r} \right) / f^2 . \label{eq:Grr}
\end{align}

Relationships between the EMT components in the $(t,r)$ and the $(v,r)$ coordinates and the corresponding forms of the Einstein equations are given in Appendix \ref{A:eq}. We use the singular nature of the Schwarzschild coordinates at the apparent horizon to extract information about the EMT.
To ensure the finite values of the curvature scalars, it is sufficient to work with
\begin{align}
     \mathrm{T} \defeq ( {\tau}{^r} -  {\tau}{_t}) / f , \quad \mathfrak{T} \defeq \big( ( {\tau}{^r})^2 + ( {\tau}{_t})^2 - 2 ( {\tau}{_t^r})^2 \big) / f^2 ,
     \label{eq:TwoScalars}
\end{align}
where the contribution of $T^\theta_{~\theta}\equiv T^\phi_{~\phi}$ is disregarded, and then to verify that the resulting metric functions do not introduce further divergences \cite{MMT:22}.

  Thus, the three effective EMT components either diverge, converge to finite limits, or converge to zero in such a way that the above combinations are finite. One   option is the scaling
\begin{align}
	{\tau}{_t} \sim f^{k_E}, \qquad {\tau}{^r} \sim f^{k_P}, \qquad {\tau}{_t^r} \sim f^{k_\Phi} , \label{tauS}
\end{align}
for some powers $k_a>1$, $a=E,P,\Phi$. Another involves convergence or divergence with the same $k\leqslant 1$. For PBHs, only solutions with $k=0,1$ are relevant.

The $k=0$ solution leads to the leading terms of the metric functions
\begin{align}
		C &= r_\sg - 4 \sqrt{\pi} r_\sg^{3/2} \Upsilon \sqrt{x} + \mathcal{O}(x) , \label{eq:k0C} \\
		h &= - \frac{1}{2}\ln{\frac{x}{\xi}} + \mathcal{O}(\sqrt{x}) , \label{eq:k0h}
\end{align}
where $\xi(t)$ is determined by choice of the time variable, and the higher-order terms are matched with the higher-order terms in the EMT expansion \cite{BMT:19,MMT:22}. The consistency condition that is given by the Einstein equation~(\ref{eq:Gtr}) results in the relationship
\begin{align}
	r'_\sg/\sqrt{\xi} =   4 \epsilon_\pm\sqrt{\pi r_\sg} \, \Upsilon , \label{eq:k0rp}
\end{align}
where $\epsilon_\pm = \pm 1$ corresponds to the expansion and contraction of the Schwarzschild sphere, respectively. The contracting Schwarzschild sphere that allows for a regular description in the $(v,r)$ coordinates corresponds to a black hole of diminishing mass. The case $r_\sg'>0$  allows for a regular description in the $(u,r)$ coordinates, where $u$ is the retarded null coordinate and corresponds to an expanding white hole. In the $(v,r)$ coordinates, the black hole metric is described by
\begin{align}
	C_+(v,r) &= r_+(v)+w_1(v)y+\cO(y^2), \label{Cpl} \\
	h_+(v,r) &= \chi_1(v) y+\cO(y^2), \label{hpl}
\end{align}
where  we used the freedom of redefining the $v$ coordinate to set $h_{+}(v,r_+) \equiv 0$ and $w_1\leqslant 1$ and $\chi_1(v)$ are some functions.  The condition $w_1\leqslant1$ is due to the definition of the Schwarzschild radius, and as we see below, this inequality is strict for the bulk of the black hole evolution. Some relationships with the quantities in the $(t,r)$ coordinates are summarized  in Appendix \ref{A:rel}, and more details are given in Ref.~\cite{MMT:22}.

Two features of this solution are to be noted: the Schwarzschild radius is a timelike hypersurface in both cases, while the NEC is violated in its vicinity. We discuss the latter issue more in Sec.~\ref{emt}. Analysis of the EMT and the metric around the smaller  root $r_\rin(t)$ of $f(t,r)=0$ leads to similar expressions. The inner horizon is also a timelike hypersurface, but the NEC is satisfied in its vicinity.

The $k=1$ solutions play a role in describing the formation of a physical black hole. The detailed properties of these solutions can be found in \cite{MT:21b,MMT:22}. Here we stress the features that are needed for the understanding of Fig.~\ref{fig:rbh}(b).

 Let the first marginally trapped surface be denoted by $r_\sg(t_\rf)$. In $(v,r)$ coordinates, it appears at some $v_\mathrm{f}$ at the circumferential radius $r_+(v_\rf)$ that corresponds to Bob's $(t_\rf, r_\sg(t_\rf)=r_+)$. For $v\leqslant \vS$, the MS mass can be expanded as
\begin{align}
	C(v,r) = \Delta(v) + r_*(v) + \sum_{i \geqslant 1} w_i(v) (r-r_*)^i ,
\end{align}
where $r_*(v)$ corresponds to the maximum of $ \mathbbm{D}_v(r)\defeq C(v,r) - r$, and the deficit function $\Delta(v)\defeq \mathbbm{D}_v(r_*)$. At the advanced time $\vS$ the location of the maximum corresponds to the first marginally trapped surface $r_*(\vS) = r_+(\vS)$ and $\Delta(\vS)=0$. For $v \geqslant \vS$, the MS mass $\Delta(v)\equiv 0$.

For $v \leqslant \vS$, we have $w_1(v) - 1 \equiv 0$ since the (local) maximum of $\mathbbm{D}_v$ is determined by $d\mathbbm{D}_v/dr=0$. For $v>\vS$ evaporation means
$r'_+(\vS) \leqslant 0$. Since the trapped region is of finite size for $v>\vS$, the maximum of $C(v,r)$ does not coincide with $r_+(v)$. As a result, $w_1(v)<1$ for $v>\vS$.
The NEC is violated in some vicinity of the apparent horizon, but not at $r=r_\sg(t_\rf)$ itself, allowing a consistent matching of the NEC-violating and the NEC-satisfying regions.

The self-consistent approach on its own cannot predict the final state of the collapsing matter. If the ultracompact object in question is a regular black hole, then there is also the final event, named the disappearance of the trapped region  $(v_\mathrm{d}, r_\mathrm{d})$  for which $w_1(v_\mathrm{d})=1$. As both the inner and outer horizon components are timelike, their intersection (or intersections) cannot join smoothly in any coordinate system, providing the coordinate-independent characterization of these events.

We can now present several important features of PBHs, particularly their application to modeling regular black holes. Recall that, in spherical symmetry, the apparent horizon (and thus the notion of a trapped region) are coordinate independent in all foliations that respect this symmetry. A general coordinate-independent notion of the (future) outer and inner horizons is introduced via the condition on Lie derivatives of the expansion of the outgoing null geodesics~\cite{HW:94}. If $\vartheta_{(l)}$ and $\vartheta_{(n)}$ are the expansions of the future-directed outgoing and
ingoing null geodesic congruences, respectively, then $\theta_{(l)}=0$ defines the apparent horizon. Its components are the {outer}   (trapping) horizon that satisfies
\be
\mathcal{L}_n\vartheta_{(l)}=n^\mu\pad_\mu\vartheta_{(l)}<0,   \label{in-out}
\ee
and the inner (trapping) horizon that satisfies
\be
\mathcal{L}_n\vartheta_{(l)}>0.
\ee

A generic representation of a RBH in Fig.~\ref{fig:rbh}(a) is distinct from the PBH-based models in several important respects. The outer horizon as defined invariantly via Eq.~\eqref{in-out} coincides with the larger root of $f(v,r)=0$. However, the NEC is violated only along section (32) and section (24) of the inner horizon. The roots of $f(u,r)=0$ do not agree with the invariant definition. The inner and the outer horizon segments join smoothly, and this is effected by having spacelike segments of both. This smooth joining prevents the identification of invariant events of formation and collapse (or evaporation) of the trapped region. This makes such models unsuitable for representing RBHs that, among other things, have a finite lifetime according to a distant observer.

Hypersurfaces of constant $r$ are timelike outside the trapped region and spacelike inside, while the opposite is true for hypersurfaces of constant $t$. We illustrate these transitions on the hypersurfaces $\Sigma_t$. A hypersurface  can be defined by restricting the coordinates via $\Psi(\Sigma_{t_0}) \eqdef t - t_0 \equiv 0$. Then $\mathfrak{l}_\mu\defeq\Psi_{,\mu}$ is the normal vector field \cite{P:04}, which is timelike for a spacelike segment of the hypersurface and spacelike for a timelike segment. Using $\Psi_{,\mu}$, one can define a normalized vector field that points in the direction of increasing $\Psi$.

Using either $(t,r)$ or $(v,r)$ coordinates, we find that \cite{DMT:22}
\begin{align}
	\mathfrak{l}_\mu \mathfrak{l}^\mu = -e^{-2h} f^{-1}.
\end{align}
   As $r\to r_\sg$ (and similarly at the inner apparent horizon), $\mathfrak{l}^2\to 0$. Thus, along $\Sigma_{t_0}$ that passes through a PBH, the normal field changes continuously. Moreover \cite{DMT:22}, at $\big(t,r_\sg(t)\big)$, the vector $\mathfrak{l}^\mu$ is proportional to $l^\mu$ of Eq.~\eqref{null-v}. Figure~\ref{fig:rbh}(b) shows the hypersurface $\Sigma_{t_\rf}$ that corresponds to the formation time of the trapped region, according to Bob. It is spacelike everywhere apart from $\big(t_\rf, r_\sg(t_\rf)\big)$ where it is null.

An ostensibly innocent requirement of finite formation time, according to Bob, has far-reaching consequences. In this case both $r_\sg(t)$ and $r_\rin(t)$ are timelike; they correspond to the invariant definitions of the outer and inner horizons, as we now show.

   A direct calculation shows that the inner and outer horizons [defined as the roots of $f(v,r)=0$, $r_{<+}(v)$ and $r_+(v)$, respectively], correspond to the invariant definition of Eq.~\eqref{in-out}. The lines of constant $u$ intersect each of these two segments only once.  If we parametrize a future-directed outgoing radial  null geodesic as $(v(\lambda),r(\lambda),0,0)$, then
   \be
   \frac{dr}{dv}=\frac{1}{2}e^{h_+(v,r)}f(v,r). \label{out-g}
   \ee
Let us assume that this geodesic intersects $r_+(v)$ twice, corresponding to the values of the affine parameter $\lambda_1<\lambda_2$. Then
\begin{align}
& v_1=v(\lambda_1), \qquad r_+(v_1)=r(\lambda_1)\eqdef r_1 \\
& v_2=v(\lambda_2), \qquad r_+(v_2)=r(\lambda_2)\eqdef r_2,
\end{align}
   and $v_1<v_2$, while $r_1>r_2$ [as $r_+'(v)<0$].  Hence, Eq.~\eqref{out-g} requires the outgoing null geodesic to pass through the trapped region $f<0$ for the values of the affine parameter $\lambda_1<\lambda<\lambda_2$. However, at the apparent horizon $dr/dv|_{r_+}=0$, making it impossible for the geodesic to enter the contracting trapped region, at least for some $\lambda>\lambda_1$.

As both the inner and the outer components of the apparent horizon are nonspacelike, they do not join smoothly, and the invariance of this taxonomy allows one to introduce well-defined events of formation and disappearance of the trapped region.

\section{EMT near the apparent horizon}\label{emt}

In spherical symmetry $ {T}^\theta_\theta \equiv {T}^\varphi_\varphi$, and the most general form of the EMT \cite{HE:73,MMT:22} in an orthonormal basis attached to a fiducial static observer is given by
\begin{align}
	{T}_{\hat\mu\hat\nu} = \begin{pmatrix}
		\rho$ $\,~ & \psi$ $~ & 0$ $~ & 0$ $~ \vspace{1mm}\\
		\psi & p & 0 & 0 \vspace{1mm}\\
		0 & 0 & { \mathfrak{p}} & 0 \vspace{1mm} \\
		0 & 0 & 0 & { \mathfrak{p}} \vspace{1mm} \\
	\end{pmatrix},
	\label{tspher}
\end{align}
where $\rho$, $p$, $\psi$, and $\mathfrak{p}$ are functions of $t$ and $r$. For $k=0$ solutions, components in the $(tr)$ block are sums of the divergent
\begin{align}
	q=-\frac{\Upsilon}{4\sqrt{\pi r_\sg x}}\, \label{eqq}
\end{align}
and additional finite terms $\mu_i$ that depend on the higher-order coefficients,
\be
\rho=q+\mu_1, \qquad p= q+\mu_3, \qquad \psi=q+\mu_2.
\ee
Classification of the EMT according to the Segre-Hawking-Ellis  scheme \cite{HE:73,MV:17} is based on the properties of the Lorentz-invariant eigenvalues of ${T}_{\hat\mu\hat\nu}$. Among the classes I--IV, the known classical matter distributions correspond to classes I and II. Type IV is considered to be the most exotic, as two of its Lorentz-invariant eigenvalues are complex conjugates. Calculations with fields that propagate on a given background are of type IV for \cite{LO:16,L:17,APT:19} $r \leqslant 1.39 r_\sg$. However, once backreaction is included, in many interesting scenarios, the more exotic forms of the EMT (types III and IV) are excluded \cite{MV:21}.

\begin{figure*}[!htbp]
  \centering
  \begin{tabular}{@{\hspace*{0.025\linewidth}}p{0.45\linewidth}@{\hspace*{0.05\linewidth}}p{0.45\linewidth}@{}}
  	\centering
   	\subfigimg[scale=0.8]{(a)}{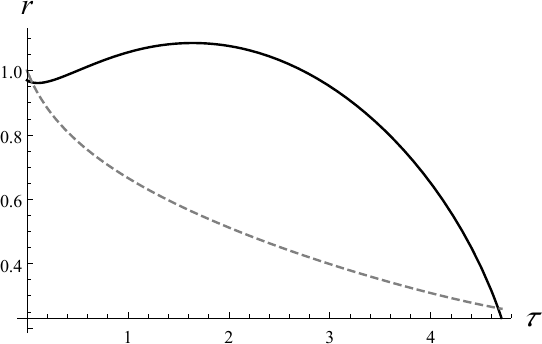} &
   	\subfigimg[scale=0.8]{(b)}{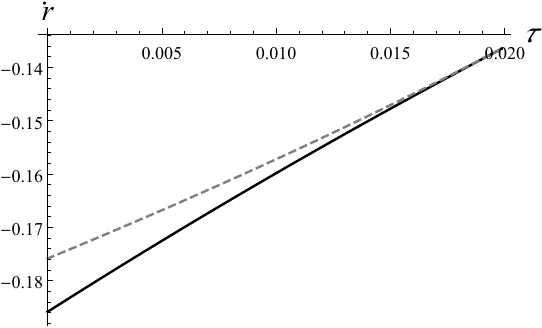}
  \end{tabular}
  	\caption{ Exit of a massive test particle from the Vaidya black hole (with the subsequent reentry). Both figures are based on the linear evaporation law $r_+(v)=r_+(0)-\alpha v$ with $r_+(0)=1$ and $\alpha=0.1$ The initial conditions are $v(0)=0$, $r(0)=0.9$ and $\dot r(0)=-\sqrt{-f\big(0,r(0)\big)}-0.01=-0.18586$.  (a) Trajectory from the initial moment until the reentry. The areal radius $r(\tau)$ (black line) and the (outer) apparent horizon $r_{+}\big(v(\tau)\big)$ (gray dashed line) are shown as   functions of the proper time $\tau$.  (b) The first segment of the trajectory until the ``reversal'' (the geodesic switches from being outgoing to ingoing). The areal radius derivative $\dot r$ is shown as a solid line, and the limiting value $-\sqrt{-f\big(v(\tau),r(\tau)\big)}$ as a dashed gray line. }
  	\label{velR}
\end{figure*}

The two nontrivial  Lorentz-invariant  eigenvalues of the EMT of Eq.~\eqref{tspher} are given by
\begin{align}
	\mathfrak{t}_{1,2} =& \half \Big(\mu_3-\mu_1\nonumber\\
	& \pm\sqrt{(\mu_1-2\mu_2+\mu_3)(\mu_1+2\mu_2+\mu_3+q)} \Big) .
\end{align}
The direct calculation (see Appendix \ref{A:rel} for details), shows that the classification at $r=r_\sg=r_+$ depends on the sign of
\be
(\mu_1-2\mu_2+\mu_3)q\propto -\theta_r^+,
\ee
i.e. it is determined by the sign of  $\Theta_{rr}(v,r_+)=e^{h_+}\Theta^v_{\,r}$ there.

The EMT is type II at $r_+$ only if the metric is sufficiently close to Vaidya, i.e., $\mu_1-2\mu_2+\mu_3=0$.

\section{Exiting the black hole}\label{exit}

The timelike nature of the retreating outer apparent horizon of a physical black hole allows for the escape of test particles from it. Outgoing null geodesics of Fig.~\ref{fig:rbh} reach the future null infinity by crossing the outer apparent horizon. An analysis of Sec.~\ref{pbh} shows that the entire regular black hole is indeed the quantum ergosphere. Null geodesics on the Vaidya background can be reduced to a system of the first-order equations, and for a linear case $C(v)=r_0-\alpha v$, it allows an analytic solution \cite{WL:86,B:16}. Appendix~\ref{B:null} uses the methods of Ref.~\cite{B:16} to describe trajectories that start inside the trapped region.

We should note that inside the trapped region $f<0$ distinction between the two families of the future-directed radial null geodesics as ``ingoing" and ``outgoing"  is not determined by their local properties and, depending on the global spacetime structure, may be purely conventional. For a mathematical black hole, the curves $v=\mathrm{const}$, as well as the outgoing null geodesics that originate inside it, reach the singularity. 
For a RBH of Fig~\ref{fig:rbh}(b), they cross the inner apparent horizon and reach $r=0$, while the outgoing geodesics cross the outer apparent horizon and reach the null infinity.

The motion of timelike test particles is more intricate. To simplify the exposition, we describe the near-horizon geometry by the Vaidya metric with $r'_+(v)<0$. The ingoing and outgoing families inside the trapped region satisfy the timelike condition
\be
-f\dot v^2+2 \dot v\dot r=-1, \label{u2-1}
\ee
implying that the components of the four-velocity are related by
\be
\dot v=\frac{\dot r\pm\sqrt{\dot r^2+f}}{f}, \label{inf}
\ee
respectively and are identified by their null limits for $|\dot r|\to \infty$.
It is known \cite{PK:22} that the contracting outer apparent horizon can overtake the test particle, releasing it (temporarily or permanently) from the black hole. Using the Vaidya metric as an example, we highlight another property: in contrast with the null case, the outgoing timelike geodesics cannot reach the apparent horizon. Beyond some value of the proper time $\tau_*$, $r(\tau_*)<r_+\big(v(\tau_*)\big)$ [that is implicitly characterized by $\dot v=1/\sqrt{-f(v,r)}$] the integral curves of the geodesic equation can be continued only by the ingoing geodesics.

The geodesic equations for the radial timelike geodesics take the form,
\begin{align}
    & \ddot r=-\frac{r_+}{2r^2}-\frac{r'_+}{2r}\dot v^2, \label{eqr0}\\
   & \ddot v=- \frac{r_+}{2r^2}\dot v^2, \label{eom5}
\end{align}
and the first term on the right-hand side of Eq.~\eqref{eqr0} is absent for null geodesics. (Equations of motion for a general metric are given in Appendix \ref{B:gen.eq})

We now consider a simple model to illustrate some of the properties of the outgoing trajectories. The model we are going to use is that of the linearly evaporating Vaidya black hole. We study a massive particle starting its motion from the black hole's interior and following an outgoing geodesic. As we can see from Fig~\ref{velR}(a), the particle can exit the PBH, but the gravitational attraction is enough to force it back inside. This behavior is heavily dependent on the initial conditions of its motion. Some initial conditions allow particles to escape forever and reach future null infinity, others,   force  a reentry, as shown in Fig~\ref{velR}(a), while other particles will never escape and head toward the centre. In all of these motions it is important to take into account Eq~\eqref{inf} which implies that inside the PBH $\dot r\leqslant-\sqrt{-f(v,r)}$. As illustrated in Fig.~\ref{velR}(b), the outgoing geodesic approaches this value until $\dot r^2=-f$ and then becomes the ingoing geodesic. So a particle can only exit the PBH following an ingoing geodesic and not an outgoing one. This behavior can be revealed by the following simple steps.

We first note that according to Eq.~\eqref{eom5}   $\dot v$ is a decreasing function of $\tau$ and thus $\dot{v}(\tau_1)>\dot{v}(\tau_2)$ for $\tau_2>\tau_1$. On the other hand, as both $\dot r$ and $f$ are negative inside the trapped region, for the outgoing geodesics, the inequality
\be
\dot v = \frac{\dot r \beta}{f}\geqslant \frac{1}{\sqrt{-f}}, \qquad 1\leqslant\beta<2, \label{vdbound}
\ee
must hold. The minimum value of $\beta=1$ corresponds to the point where $\dot r=-\sqrt{-f}$.

Thus, the outgoing geodesics [that starts at some $r(0)<r_+(0)$ with some finite value of $\dot{v}(0)$], cannot exit through $r_+$, as in this case $f(v,r=r_+)=0$ implies the divergence of $\dot v$ via Eq.~\eqref{vdbound}. However, as it is a decreasing function according to Eq.~\eqref{eom5}, such occurrence is impossible, and there should be a value $\tau_*$ where the geodesic changes from being outgoing to being ingoing. It occurs continuously at the point where
\be
\dot r(\tau_*)=-\sqrt{-f\big(v(\tau_*),r(\tau_*)\big)}, \label{turn}
\ee
and $\dot v=1/\sqrt{-f}$.

\section{Tidal forces}\label{tidal}

Curvature scalars are finite by construction at the apparent horizon of a PBH. However, the absence of these singularities (so-called parallel propagated or p.p. singularities, \cite{HE:73}) does not rule out a weaker form of singular behavior. In fact, the apparent horizon is a weakly singular surface\cite{DMT:22,MMT:22}. For example,  it is possible to find a null tetrad where one of the Ricci spinors $\Phi_{00}$ or $\Phi_{22}$ diverge. The energy density is finite from the point of view of an infalling observer, geodesic or not, so long as $\dot r<0$. On the other hand,  the energy density diverges in the proper frame of an outgoing test particle on a nongeodesic trajectory that approaches the outer apparent horizon.

Consider a trajectory $x^\mu_\mA(\tau)$ that is implicitly given for its entire duration inside the black hole by Eq.~\eqref{turn}. Curiously enough the geodesic equation~\eqref{eqr0} is satisfied, even if Eq.~\eqref{eom5} is not.
In this case, close to the apparent horizon, the energy density is
\be
\rho_\mA=T_{\mu\nu} \dot x^\mu_\mA\dot x^\nu_\mA\approx \frac{r'_+}{8\pi r_+|y|},
\ee
where $y=r-r_+$. It diverges, but the integrated energy density remains finite. The divergence has an intuitive explanation if one notes that the square of the four-acceleration diverges as $r\to r_+$.

Divergent tidal forces are one of the hallmarks of spacetime singularities. It is a standard textbook result \cite{MTW:73} that the tidal forces on an infalling Alice at the horizon of a Schwarzschild black hole are large but finite. Falling through the apparent horizon of the Vaidya black hole is qualitatively similar. Indeed,  using the geodesic deviation equation to determine the three components of acceleration in the proper frame of Alice with $\dot r=1$ at the apparent horizon (see Appendix~\ref{b:tidal} for the detailed description of the frame),
\be
\frac{D^2\zeta^{(j)}}{d\tau}=-R_{(\tau)(j) (\tau)(k)}\zeta^{(k)},
\ee
where $j,k=\rho,\theta, \phi$ the three nonzero curvature terms are
\begin{align}
R_{(\tau)(\rho)(\tau)(\rho)}=&-\frac{r_+(v)}{r^3}=-\frac{1}{r_+^2}+\cO(y),\\
R_{(\tau)(\theta)(\tau)(\theta)}=&\frac{1}{2 r_+^2}\left(1+\frac{r'_+}{4}\right)+\cO(y),\\
\end{align}
where $y=r-r_+$ and $R_{(\tau)(\phi)(\tau)(\phi)}=R_{(\tau)(\theta)(\tau)(\theta)}$. Thus, evaporation produces just corrections that are proportional to the evaporation rate to the tidal force experienced by the infalling Alice.

The result is the same if the retreating apparent horizon overtakes the infalling particle inside the RBH. The situation is different for a nongeodesic outgoing particle. Consider again the trajectory that is implicitly given for its entire duration inside the black hole by Eq.~\eqref{turn}. Then the two nonradial tidal force components diverge as
\be
R_{(\tau)(\theta)(\tau)(\theta)}=-\frac{r'_+}{2r_+y}+\cO(y^0).
\ee

\section{Discussion}

We studied the properties a spherically symmetric PBH must have in the context of semiclassical gravity. An ostensibly innocent requirement of finite formation time, according to Bob, has far-reaching consequences. The NEC must be violated in the vicinity of the outer apparent horizon, but it is satisfied in the vicinity of the inner horizon. We have well defined events of formation and disappearance of the trapped region because both horizons are timelike (Fig.\ref{fig:rbh}). The EMT classification, according to the Serge-Hawking-Elis scheme on the apparent horizon is of type I, under certain assumptions, consistent with what is believed to happen when backreaction is included.

In contrast with the Schwarzschild solution, both massless and massive particles inside the quantum ergosphere of a PBH are able to escape the trapped region. For massless particles, this is evident from the timelike character of the outer apparent horizon. A careful analysis of massive particles' trajectories, using the Vaidya limit as an example, shows that they can only escape when following ingoing geodesic trajectories. We calculated the tidal forces experienced by observers, in general, for objects of finite size. For infalling geodesic observers, the tidal forces are finite. This is also the case for observers inside the quantum ergosphere, since they can only cross the outer apparent horizon when following an ingoing geodesic and letting the receding horizon overtake them. This is not the case for nongeodesic observers who experience infinite tidal forces when they try to force themselves out of the trapped region. Furthermore, nongeodesic observers experience infinite negative energy density in the form of a firewall \cite{DMT:22} when they try to escape the trapped region, something that does not happen for geodesic observers. Despite the fact that the apparent horizon is assumed to be regular in the sense of finite curvature scalars, all these properties indicate that it possesses a mildly singular nature manifesting itself as infinite tidal forces and firewalls for specific observers.
Astrophysical black holes are rotating, and study of general axially symmetric PBHs is subject of our future research. However,   a special case of the Kerr-Vaidya metric illustrates that the violation of the NEC and  a mild firewall are not  artifacts of spherical symmetry~\cite{dp31}. In the Kerr-Vaidya geometry, which is of Petrov-II, the NEC is always violated due to the type III EMT (based on the Serge-Hawking-Ellis classification \cite{HE:73,MV:17}) on the apparent horizon. Moreover, in the equatorial plane of the Kerr-Vaidya metric there are radial geodesics  whose equations of motion are the same as those of their counterparts in the Vaidya metric. Thus, while we expect that the axial symmetry introduces more complicated scenarios of motion of test particles, they also include the results that were described above.

\section*{Acknowledgements}
Useful discussions with Eleni Kontou, Sebastian Murk, Joe Schindler and Justin Tzou are gratefully acknowledged. P.K.D. and I.S. are supported by an International Macquarie University Research Excellence Scholarship. The work of D.R.T. is supported by the ARC Discovery project Grant No. DP210101279.
\appendix

\section{SUMMARY OF USEFUL RELATIONS}
\subsection{Einstein equations and basic definitions}\label{A:eq}
A useful relationship between the EMT components in $(t,r)$ and $(v,r)$ coordinates is given by
\begin{align}
	&	 {\theta}{_v} \defeq e^{-2h_+}  {\Theta}_{vv} =  {\tau}{_t} , \label{eq:thev} \\
	&	 {\theta}_{vr} \defeq e^{-h_+}  {\Theta}_{vr} = \left(  {\tau}{_t^r} -  {\tau}{_t} \right) / f , \label{eq:thevr}\\
	&	 {\theta}{_r} \defeq  {\Theta}_{rr} = \left(  {\tau}{^r} +  {\tau}{_t} - 2  {\tau}{_t^r} \right) / f^2 ,   \label{eq:ther}
\end{align}
where $ {\Theta}_{\mu\nu}$ denotes the EMT components in $(v,r)$ coordinates. We denote the limit of $ {\theta}{_v}$ as $r \to r_+$ as $\theta^+_v$, etc. The Einstein equations are then given by
\begin{align}
	&  \pad_v C_+  = 8 \pi r^2 e^{h_+}( {\theta}{_v}+ f  {\theta}_{vr})\equiv 8 \pi r^2  \Theta^r_{\;v} , \label{eq:Gvv}\\
	& \pad_r C_+ = - 8 \pi r^2  {\theta}_{vr} \equiv 8\pi r^2 \Theta^v_{\;v}, \\
	& \pad_r h_+ = 4 \pi r  {\theta}{_r}\equiv  4 \pi r e^{h_+}\Theta^v_{\;r}  . \label{vGrr}
\end{align}
Tangents to the congruences of ingoing and outgoing radial null geodesics (note that these designations make  literal sense only in a space with simple topology) are given in $(v,r)$ coordinates by
\begin{align}
	n^\mu=(0,-e^{-h_+},0,0), \qquad l^\mu=(1,\half e^{h_+}f,0,0), \label{null-v}
\end{align}
respectively.   The vectors are normalized to satisfy $n\cdot l=-1$. Their expansions \cite{HE:73,P:04,C-B:09} are
\begin{align}
	\vartheta_{(n)} = - \frac{2e^{-h_+}}{r}, \qquad \vartheta_{(l)}=\frac{e^{h_+}f}{r},
\end{align}
respectively. Hence the (outer) apparent horizon is located at the Schwarzschild radius $r_\sg$  \cite{vF:15,B:16,FEFHM:17}, justifying the definition of the black hole mass as \cite{jB:81} $2M(v)=r_+(v)$.

\subsection{Velocity components, EMT and metric functions} \label{A:rel}

In $(v,r)$ coordinates outside of the apparent horizon the relationship between four-velocity components of the timelike trajectory is
\begin{align}
	\dot{v}=\frac{\dot r+\sqrt{\dot r^2+f}}{e^{h_{+}} f},\label{vdin}
\end{align}
for both ingoing ($\dot r<0$) and outgoing ($\dot r>0$) test particles, where $f=f\left(v(\tau),r(\tau)\right)$, and $h_{+}=h_+\left(v(\tau),r(\tau)\right)$. On the other hand, inside  the trapped region $f<0$ and thus to maintain the timelike character of the trajectory
\begin{align}
	\dot r\leqslant-\sqrt{ -f}
\end{align}
must hold.
The null velocity component  of ingoing particles still satisfies Eq.~\eqref{vdin}, with the ingoing null geodesics $\dot v=0$ being their ultrarelativistic limit. The future-directed outgoing trajectories satisfy
\begin{align}
	\dot{v}=\frac{\dot r-\sqrt{\dot r^2+f}}{e^{h_{+}} f}>0. \label{voutA}
\end{align}
The limiting values of the EMT components in $(v,r)$ coordinates, $\theta^+_{\mu\nu} \defeq \lim_{r\to r_+} \Theta_{\mu\nu}$, are
\begin{align}
	& \theta_v^+= (1-w_1)\frac{r_+'}{8\pi r_+^2}, \label{the1} \\
  	& \theta_{vr}^+=- \frac{w_1}{8\pi r_+^2}, \label{the2} \\
  	& \theta_r^+=\frac{\chi_1}{4\pi r_+}. \label{the3}
\end{align}

The effective EMT components in the $(t,r)$ coordinates for $x\defeq r-r_\sg(t)>0$ are
\begin{align}
	&  {\tau}{_t} = - \Upsilon^2 + e_{12}\sqrt{x} + e_1 x + \cO(x^{3/2}) , \label{eq:k0taut} \\
	&  {\tau}{_t^r} = \pm \Upsilon^2 + \phi_{12} \sqrt{x} + \phi_1 x + \cO(x^{3/2}) , \label{eq:k0tautr} \\
	&  {\tau}{^r} = - \Upsilon^2 + p_{12} \sqrt{x} + p_1 x + \cO(x^{3/2}) . \label{eq:k0taur}
\end{align}
Since $f\propto \sqrt{x}$ and the rhs of Eq.~\eqref{eq:Grr} results in a finite limit, we have
\begin{align}
	\phi_{12}=\half (e_{12} + p_{12}).
\end{align}
The metric functions are then given by
\begin{align}
C &=r_\sg- 4 \sqrt{\pi r^{3}_\sg} \Upsilon \sqrt{x}+\left(\frac{1}{3}+\frac{4 \sqrt{\pi} e_{12}r_\sg^{3/2}}{3\Upsilon}\right) x + \cO(x^{3/2}),\\
h&= -\frac{1}{2}\ln\frac{x}{\xi}+\left(\frac{1}{3\sqrt{\pi}r_\sg^{3/2}\Upsilon}-\frac{e_{12} - 3 p_{12}}{6\Upsilon^2}\right)\sqrt{x}+\cO(x).
\end{align}
By substitution into Eq.~\eqref{eq:thevr}, we find that the limiting values of the effective EMT components in $(v,r)$ coordinates are given by
\begin{align}
	\theta_v^+ & = - \Upsilon^2, \label{thevUps}\\
	\theta_{vr}^+ &= \frac{p_{12} - e_{12}}{8\sqrt{\pi r_\sg}\Upsilon}, \label{thevrA} \\
	\theta_r^+ &= \frac{e_1 - 2 \phi_1 + p_1}{16\pi r_\sg\Upsilon^2}. \label{therrA}
\end{align}
The leading terms of the EMT components of Eq.~\eqref{tspher}  are expressed with 
\begin{align}
\mu_1&=\frac{4\sqrt{\pi} e_{12} r_\sg^{3/2}+\Upsilon}{24\pi r_\sg^2\Upsilon} +\cO(\sqrt{x}),\\
\mu_3&=\frac{ \sqrt{\pi}r_\sg^{3/2}(e_{12} + 3 p_{12})+\Upsilon}{24\pi r_\sg^2\Upsilon}+\cO(\sqrt{x}),\\
\mu_2&=\half(\mu_1+ \mu_3)+\cO(\sqrt{x}).
\end{align}

The quantity
\be
\mu_1-2\mu_2+\mu_3=\frac{(e_1-2\phi_1+p_1)\sqrt{x}}{4\sqrt{\pi r_\sg}\Upsilon}+\cO(x)
\ee
is important  for the EMT classification. Using Eqs.~\eqref{therrA} and \eqref{eqq} we find
\be
(\mu_1-2\mu_2+\mu_3)q=-\frac{(e_1-2\phi_1+p_1)}{16 {\pi r_\sg}}\propto-\theta_r^+,
\ee
and due to Eq.~\eqref{vGrr} $\Theta_{rr}\propto\Theta^v_{\;r}$.

From Eqs.~\eqref{eq:Gvv} and \eqref{thevUps} it follows that $\Theta^r_{\;v}<0$ at the apparent horizon.
 If at the outer apparent horizon $\Theta^v_{\;r}\sim\Theta^r_{\;v}$, then the EMT has only real Lorentz-invariant eigenvalues and belongs to type I.

\section{EXITING RBH}\label{B:ap}

\subsection{Equations of motion on the background of a general spherically symmetric metric }\label{B:gen.eq}
For the general spherically symmetric metric in the $(v,r)$ coordinates
\begin{align}
	ds^2=-e^{2h_+}f(v,r)dv^2+2e^{h_+}dvdr +r^2d\Omega_2,
\end{align}
we have the following equations of motion for radially moving massive particles
\begin{align}
    & \ddot{v}+\left((\partial_{v}h_{+})+e^{h_{+}}(\partial_{r}h_{+})+\frac{1}{2}e^{h_{+}}(\partial_{r}f)\right) \dot{v}^2=0, \label{bddv}\\
    & \ddot{r}+(\partial_{r}h_{+})\dot{r}^2+\left(-\frac{1}{2}e^{h_{+}}(\partial_v f)\right)\dot{v}^2,\nonumber\\
    & +\left( (\partial_rh_{+})f+\frac{1}{2}(\partial_rf)\right)=0,
 \end{align}
with the timelike normalization condition
\begin{align}
     -e^{2h_{+}}f\dot{v}^2+2e^{h_{+}}\dot{v}\dot{r}=-1.
\end{align}
The above relations for $h_{+}=0$ and $f(v,r)=1-r_{+}(v)/r$ reduce to the geodesic equations~\eqref{eqr0} and~\eqref{eom5} for the Vaidya metric.

\subsection{Outgoing null geodesics in the ingoing Vaidya metric}\label{B:null}

Here we consider the future-directed null geodesics that are outgoing from the Vaidya black hole with $r'_+(v)<0$. They satisfy the geodesic equation
\be
 \dot v=0, \qquad \dot v=\frac{2\dot r}{f},\label{norm0}
\ee
for the ingoing and outgoing geodesics.

Adapting the results of Refs.~\cite{WL:86,B:16}, the equations are simplified by introducing $\dot r_+=r'_+\dot v$ and noting
\be
\ddot r+\frac{r'_+}{2r}\dot v^2=\ddot r+\frac{\dot r\dot r_+}{r-r_+}=\ddot r-\dot r\frac{d}{d\lambda}\ln(r-r_+)+\frac{\dot r^2}{r-r_+}.
\ee
For a nonzero $\dot r$ (that is definitely true up to the apparent horizon), it can be rearranged as
\be
\frac{d}{d\lambda}\ln\frac{\dot r}{r-r_+}=-\frac{\dot r}{r-r_+},
\ee
whose integration results in
\be
\frac{\dot r}{r-r_+}=\frac{1}{\lambda+c},
\ee
where $c$ is set by the initial conditions. Hence Eq.~\eqref{eqr0} is equivalent to
\be
\dot r(\lambda)=\frac{r(\lambda)-r_+\big(v(\lambda)\big)}{\lambda+c}.
\ee
Then Eq.~\eqref{norm0} results in
\be
\dot v(\lambda)=\frac{2 r(\lambda)}{\lambda+c}.
\ee
The initial condition [we can choose $\lambda=0$, $v(0)=0$] identifies the constant as
\be
c=\frac{r(0)-r_0}{\dot r(0)}>0.
\ee

If the black hole evaporation rate is constant, $r_+=r_0-\alpha v$, for some $\alpha>0$, then the substitution
\be
\lambda+c=c  e^{\ell}
\ee
results in a first-order linear system
\begin{align}
\frac{dr}{d\ell}&=r(\ell)-r_0+\alpha v(\ell), \\
\frac{dv}{d\ell}&=2r(\ell),
\end{align}
which can be rewritten in the form of matrices as
\begin{equation}
    \frac{d}{d\ell} \begin{pmatrix}
    r\\
    v
    \end{pmatrix}=
    \begin{pmatrix}
1 & \alpha \\
2 & 0
\end{pmatrix}
\begin{pmatrix}
    r\\
    v
    \end{pmatrix}+
    \begin{pmatrix}
    -r_0\\
    0
    \end{pmatrix},
\end{equation}
with the matrix
\begin{align}
    A=\begin{pmatrix}
1 & \alpha \\
2 & 0
\end{pmatrix}
\end{align}
having eigenvalues
\begin{align}
    \omega_1=\frac{1}{2}\left(1+\sqrt{1+8\alpha}\right)>0,
\end{align}
\begin{align}
    \omega_2=\frac{1}{2}\left(1-\sqrt{1+8\alpha}\right)<0,
\end{align}
so one can find that the solution to the linear system is
\begin{align}
 r(\ell)=\frac{1}{\omega_1-\omega_2}\big(\left(-r_0+\omega_1r(0)\right)e^{\omega_1\ell}+\left(r_0-\omega_2r(0)\right)e^{\omega_2\ell}\big), \label{pos:r}
\end{align}
\begin{multline}
    v(\ell)= -\frac{2r_0}{\omega_1\omega_2}+\frac{1}{\omega_1-\omega_2}\Bigg(\left(r(0)-\frac{r_0}{\omega_1}\right)e^{\omega_1 \ell}\\
    +\left(-r(0)+\frac{r_0}{\omega_2}\right)e^{\omega_2 \ell}\Bigg).
\end{multline}

\begin{figure}[htbp]
	\includegraphics[width=0.45\textwidth]{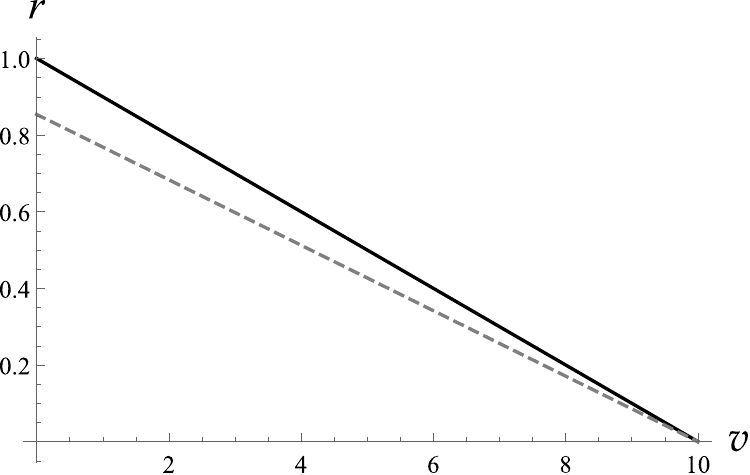}
	\caption{The black line represents the evolution of the apparent horizon $r_{+}(v)=r_0-av$, following a linear evaporation law, and the gray dashed line represents the evolution of the null hypersurface $r_{\mathrm{S}}(v)$. For this diagram, the constants $\alpha$ and $r_0$ are chosen to be $r_0=1$ and $\alpha=0.1$. }
	\label{fig:EH}
\end{figure}

The event horizon (Fig~\ref{fig:EH}) is found by identifying the rays that do not reach the future null infinity ${\mathscr{I}}^+$. As $\omega_2<0$,  the second term on Eq.~$\eqref{pos:r}$ vanishes in the limit $\ell\rightarrow +\infty$. Hence the arrival of a null particle  $\mathscr{I}^+$ depends on the sign of the coefficient of the first term of the equation of trajectory,
\be
\delta\defeq \omega_1 r(0)-r_0.
\ee
The null particles that at $\lambda=0$ start from the areal radius $r<r_\mathrm{eh}(0)$,
\begin{align}
r_\mathrm{eh}(0)=\frac{r_0}{\omega_1}=\frac{2r_{+}(0)}{1+\sqrt{1+8\alpha}},
\end{align}
do not reach $\mathscr{I}^+$. By varying  the initial affine parameter, one obtains the following time-dependent expression for this hypersurface,
\begin{align}
r_\mathrm{eh}(\ell)=\frac{2r_{+}(v(\ell))}{1+\sqrt{1+8\alpha}}.
\end{align}
It is indeed a null hypersurface, as can be easily verified by calculating the norm of the normal vector. Describing the hypersurface as
\begin{align}
    \Psi=r-r_{\mathrm{eh}}=0,
\end{align}
we find that the normal vector   $\mathfrak{l}_{\mu}=-\partial_{\mu}\Psi$ is given by
\begin{align}
 \mathfrak{l}_{\mu}=\left(-\frac{2\alpha}{1+\sqrt{1+8\alpha}},-1,0,0\right)
\end{align}
and satisfies  $\mathfrak{l}^{\mu}\mathfrak{l}_{\mu}=0$.
 It is also worth noting that for very small evaporation rate $\alpha \ll 1$, one can show that the hypersurface is approximately the separatrix \cite{BHL:18},
\begin{equation}
   r_{\mathrm{eh}}\approx r_{+}(1-2\alpha)=r_{+}+2r'_{+}r_{+}.
\end{equation}

\subsection{Turning point for the outgoing timelike geodesics }
As $h_+=0$ at the apparent horizon, Eq.~\eqref{voutA} implies that $\dot v\to +\infty$ as a test particle on the outgoing timelike trajectory approaches the apparent horizon form inside. On the other hand, $\ddot v$ is always finite, and at the apparent horizon
\be
\ddot v_{|_{r_+}}=-\left(\chi_1+\frac{1-w_1}{r_+}\right),
\ee
where we substituted the expansion of Eqs.~\eqref{Cpl} and \eqref{hpl} into the geodesic equation~\eqref{bddv}. If $\chi_1>0$ (i. e., the EMT belongs to the type I), then $\ddot v<0$ in some vicinity of the apparent horizon. In this case, the arguments of Sec.~\ref{exit} that establish the existence of the turning point apply without any change.

Even when it cannot be asserted that $\dot v$ is a decreasing function of the proper time, $\ddot v$ remains finite. As a result, $\dot v$ diverges only if the outgoing timelike geodesic reaches $r_+$   as $\tau\to\infty$. However,  it implies that $v\to\infty$, contradicting the assumption of finite evaporation time.

\section{TIDAL FORCE CALCULATIONS}\label{b:tidal}

For a radially moving observer with the four-velocity $u_\mA=(\dot v,\dot r, 0,0)$, the comoving orthogonal tetrad is formed by
\begin{align}
&e_{(\lambda)}^\mu=u^\mu_\mA, \\ &e_{(\rho)}^\mu=\frac{1}{\sqrt{{e^{2h_+}f-2 e^{h_+}\dot r/\dot v}}}\left(1, {e^{h_+}f-\dot r/\dot v},0,0\right), \\
&e_{(\theta)}^\mu=\left(0,0,1/r,0\right), \\
&e_{(\phi)}^\mu=\left(0,0,0,1/(r \sin\theta)\right).
\end{align}

For a massive test particle falling through the apparent horizon, a useful approximate expression for the four-velocity $u_\mA$ can be readily obtained. We set
\be
\dot r^2\eqdef -f(v,r)+E^2+r_+'(v)\, g(v,r) \label{dr10},
\ee
where the function $g(v,r)$ is to be determined.

For a Schwarzschild black hole, the constant $E$ represents the conserved energy per unit mass, and $g\equiv 0$. If the test particle starts from infinity with zero velocity, then $E=1$.

Expanding in powers of $y\defeq r-r_+$, we have
\be
f=y/r_+ +\cO(y^2),
\ee
and assuming
\be
g=\gamma(v)y+\cO(y^2),
\ee
for some function $\gamma(v)$, we have
\be
\dot r=-E+\left(\frac{1}{r_+}-\gamma r'_+\right)\frac{y}{2E}+\cO(y^2).
\ee

Outside the apparent horizon
\be
\dot v=\frac{\dot r+\sqrt{f+\dot r^2}}{f}, \label{vext}
\ee
for both ingoing and outgoing geodesics. Thus,
\be
\dot v= \dot v_0+\dot v_1+\cO(y^2), \label{vout}
\ee
where
\be
\dot v_0=\frac{1}{2E}, \qquad \dot v_1=\frac{y}{8E^3}\left(\frac{1}{r_+}-2\gamma r'_+\right),
\ee
and here and in the following, we keep only the leading-order terms in both $y$ and $r'_+$.

Taking the derivative over the proper time $\tau$ and substitution into the equation of motion~\eqref{eom5} results in the identity
\be
\frac{(1-4 E^2\gamma r_+)r'_+}{16 E^4 r_+}+\cO(y)=0,
\ee
from which the leading correction  coefficient
\be
\gamma=\pad_r g\big(v,r_+(v)\big)=\frac{1}{4E^2 r_+}
\ee
is extracted. As a result, the radial velocity in the vicinity of the apparent horizon is approximated as
\be
\dot r^2=E^2+\frac{y}{r_+}\left(1+\frac{r'_+ }{4E^2}\right)+\cO(y^2).
\ee
To compare the results of Sec.~\ref{tidal} with the standard estimates of the experiences of a freely falling observer at the black hole horizon \cite{MTW:73}, we set $E=1$ in the calculations of Sec.~\ref{tidal}.


\begin{thebibliography}{99}

\bibitem{HE:73} S.\ W.\ Hawking and G.\ F.\ R.\ Ellis,
	{\href{https://doi.org/10.1017/CBO9780511524646}{\textit{The Large Scale Structure of Space-Time} (Cambridge University Press, Cambridge, England, 1973)}}.
\bibitem{MTW:73}  C.~Misner, K.~Thorne, and J.~A.~Wheeler,
	\textit{Gravitation} (Princeton University Press, Princeton, NJ,1973).

\bibitem{FN:98} V.\ P.\ Frolov and I.\ D.\ Novikov,
	{\href{https://doi.org/10.1007/978-94-011-5139-9}{\textit{Black Hole Physics: Basic Concepts and New Developments} (Kluwer, Dordrecht, 1998)}}.
\bibitem{P:04} E.\ Poisson,
	{\href{https://doi.org/10.1017/CBO9780511606601}{\textit{A Relativist's Toolkit: The Mathematics of Black-Hole Mechanics} (Cambridge University Press, Cambridge, England, 2004)}}.


\bibitem{LIGO:21} LIGO Scientific Collaboration, KAGRA Collaboration, and Virgo Collaboration,
	{\href{https://doi.org/10.3847/2041-8213/ac082e}{ {Astrophys.\ J.\ Lett.} \textbf{915}, L5 (2021)}}.
\bibitem{eC:19} E. Curiel,
	{\href{https://doi.org/10.1038/s41550-018-0602-1}{ {Nat.\ Astron.} \textbf{3}, 27 (2019)}}.

\bibitem{F:14} V.~P.~Frolov,
	{\href{https://arxiv.org/abs/1411.6981}{arXiv:/1411.6981}}.

\bibitem{H:00} S. A. Hayward, 
{\href{https://arxiv.org/abs/gr-qc/0008071v2}{arXiv:gr-qc/0008071}}.

\bibitem{mV:14} M.~Visser,
	{\href{https://doi.org/10.1103/PhysRevD.90.127502}{ {Phys.\ Rev.\ D} \textbf{90}, 127502 (2014)}}.	

\bibitem{vF:15} V.~Faraoni,
 	{\href{https://doi.org/10.1007/978-3-319-19240-6}{\textit{Cosmological and Black Hole Apparent Horizons} (Springer, Heidelberg, 2015)}}.

\bibitem{MMT:22} R. B. Mann, S. Murk, and D.  R. Terno, 
\href{https://www.worldscientific.com/doi/10.1142/S0218271822300154}{Int. J. Mod. Phys. D \textbf{31}, 2230015 (2022)}.


\bibitem{BMMT:19}
V. Baccetti, R. B. Mann, S. Murk, and D.~R. Terno,  \href{https://journals.aps.org/prd/abstract/10.1103/PhysRevD.99.124014}{Phys. Rev. D \textbf{99}, 124014 (2019).}

\bibitem{MMT:22D} R.\ B.\ Mann, S.\ Murk, and D.\ R.\ Terno,
	\href{https://doi.org/10.1103/PhysRevD.105.124032}{Phys. Rev. D \textbf{105}, 124032 (2022).}

\bibitem{FEFHM:17} V.~Faraoni, G.~F.~R.~Ellis, J.~T.~Firouzjaee, A.~Helou, and I.~Musco,
	{\href{https://doi.org/10.1103/PhysRevD.95.024008}{{Phys.\ Rev.\ D} \textbf{95}, 024008 (2017)}}.

\bibitem{BMT:19} V.\ Baccetti, S.\ Murk, and D.\ R.\ Terno,
	{\href{https://doi.org/10.1103/PhysRevD.100.064054}{{Phys.\ Rev.\ D} \textbf{100}, 064054 (2019)}}.

\bibitem{MT:21b} S.\ Murk and D.\ R.\ Terno,
	{\href{https://doi.org/10.1103/PhysRevD.104.064048} {{Phys.\ Rev.\ D} \textbf{104}, 064048 (2021)}}.

\bibitem{DMT:22} P.\ K.\ Dahal, S.\ Murk, and D.\ R.\ Terno,
	{\href{https://doi.org/10.1116/5.0073598}{{AVS Quantum Sci.} \textbf{4}, 015606 (2022)}}.

\bibitem{BHL:18} P.\ Bin\'{e}truy, A.\ Helou, and F.\ Lamy,
	{\href{https://doi.org/10.1103/PhysRevD.98.064058}{ {Phys.\ Rev.\ D} \textbf{98}, 064058 (2018)}}.

   \bibitem{CDLV:20} R.\ Carballo-Rubio, F.\ Di Filippo, S.\ Liberati, and M.\ Visser,
   {\href{https://doi.org/10.1103/PhysRevD.101.084047}{ {Phys.\ Rev.\ D} \textbf{101}, 084047 (2020)}}.


\bibitem{MV:17} P.\ Mart\'{\i}n-Moruno and M.\ Visser,
	\textit{Classical and semi-classical energy conditions}, in {\href{https://doi.org/10.1007/978-3-319-55182-1}{\textit{Wormholes, Warp Drives and Energy Conditions,} edited by F.\ N.\ S.\ Lobo (Springer, New York, 2017)}}, p.\ 193.

\bibitem{LO:16} A.\ Levi and A.\ Ori,
	{\href{https://doi.org/10.1103/PhysRevLett.117.231101}{{Phys.\ Rev.\ Lett.} \textbf{117}, 231101 (2016)}}.

 \bibitem{L:17} A.\ Levi,
	{\href{https://doi.org/10.1103/PhysRevD.95.025007}{{Phys.\ Rev.\ D} \textbf{95}, 025007 (2017)}}.

 \bibitem{MV:21} P.\ Mart\'{\i}n-Moruno and M.\ Visser,
	{\href{https://doi.org/10.1103/PhysRevD.103.124003}{{Phys.\ Rev.\ D} \textbf{103}, 124003 (2021)}}.



\bibitem{APT:19} S.\ Abdolrahimi, D.\ N.\ Page, and C.\ Tzounis,
 	{\href{https://doi.org/10.1103/PhysRevD.100.124038}{{Phys.\ Rev.\ D} \textbf{100}, 124038 (2019)}}.

\bibitem{WL:86} R. Waugh, K. Lake, 
{\href{https://doi.org/10.1016/0375-9601(86)90304-X}{Phys. Lett. \textbf{116B}, 154 (1986).}}


\bibitem{B:16} M.\ Blau, \textit{Lecture Notes on General Relativity} (2016), 
	{\href{http://www.blau.itp.unibe.ch/newlecturesGR.pdf}{http://www.blau.itp.unibe.ch/newlecturesGR.pdf}}.

\bibitem{PK:22} J.\ Piesnack and K.\ Kassner,
	{\href{https://doi.org/10.1119/10.0006367}{ {Am.\ J.\ Phys.} \textbf{90}, 37 (2022)}}.

\bibitem{HW:94} S. A. Hayward,
	{\href{https://doi.org/10.1103/PhysRevD.49.6467}{{Phys.\ Rev.\ D} \textbf{49}, 6467 (1994)}}.
	
\bibitem{C-B:09} Y. Choquet-Bruhat,
	 {\href{https://academic.oup.com/book/311}{\textit{General Relativity and the Einstein Equations} (Oxford University Press, Oxford, England, 2009).}}
	
\bibitem{jB:81} J. M. Bardeen,
         \href{https://doi.org/10.1103/PhysRevLett.46.382}{Phys. Rev. Lett. \textbf{46}, 382 (1981).}

\bibitem{dp31}
P.~K.~Dahal and D.~R.~Terno,
\href{https://doi.org/10.1103/PhysRevD.102.124032}{Phys. Rev. D \textbf{102}, 124032 (2020).}

\end{thebibliography}
\end{document}